# Parameter Estimates for a Polymer Electrolyte Membrane Fuel Cell Cathode

Qingzhi Guo,[*] Vijay A. Sethuraman,[*] and Ralph E. White [**z]

Center for Electrochemical Engineering, Department of Chemical Engineering,

University of South Carolina, Columbia, South Carolina 29208, USA

[*]      Electrochemical Society Student Member

[**]     Electrochemical Society Fellow

[z]      Correspondence: White@engr.sc.edu

# Abstract

Five parameters of a model of a polymer electrolyte membrane fuel cell cathode (the porosity of the gas diffusion layer, the porosity of the catalyst layer, the exchange current density of the oxygen reduction reaction, the effective ionic conductivity of the electrolyte, and the ratio of the effective diffusion coefficient of oxygen in a flooded spherical agglomerate particle to the squared particle radius) were determined by the least square fitting of experimental polarization curves.

Key words: nonlinear parameter estimation, sensitivity approach, polarization curve, air cathode, polymer electrolyte membrane fuel cell.

# Introduction

The air cathode in a polymer electrolyte membrane fuel cell (PEMFC) is the largest source of voltage loss due to limitations of ionic (proton) conduction, multi-component gas diffusion, and liquid phase $O_2$ diffusion.[1-3] To obtain a better understanding of these limitations, several models have been presented.[1-8] Two different pictures of the catalyst layer (CAL) have been used to model the steady state polarization performance of a PEMFC cathode: the flooded CAL and the CAL with the existence of gas pores. The assumption of a flooded CAL was found to over estimate the product of the diffusion coefficient and the concentration of $O_2$ in the liquid electrolyte,[1] whereas a steady state polarization model including gas pores in the CAL was found to be more realistic.[3,5,8]

The objective of this work was to use our previously submitted air cathode model [8] that includes gas pores in the CAL to estimate the values of the GDL porosity, the CAL porosity, the exchange current density of the $O_2$ reduction reaction, the effective ionic conductivity of the electrolyte and the ratio of the effective diffusion coefficient of $O_2$ in a flooded spherical agglomerate particle to the squared particle radius from the experimental steady state polarization curves of the cathode of an air/$H_2$ PEMFC by the least square fitting. Due to the fact that the air cathode is the most important source of voltage loss in a PEMFC and the voltage loss on the $H_2$ anode is negligible, the experimental polarization curves of a PEMFC air cathode can be obtained from those of a full PEMFC after correcting for the voltage drop across the PEM.[1,7] In general, the model used here is similar to a model described in Jaouen *et al.*'s work.[3] The CAL is assumed to consist of many flooded spherical agglomerate particles surrounded by gas pores. As shown in Fig. 1, $O_2$ gas diffuses through gas pores in both the GDL and the CAL first, then dissolves into liquid water on the surface of the flooded agglomerate particles, and finally diffuses to the Pt catalyst sites or carbon surface. Protons are supplied to the Pt catalyst sites via the hydrated Nafion ionomer network in the flooded agglomerate particles. As concluded in ref. 8, it is in the liquid form that the generated water (by the $O_2$ reduction reaction) is removed out of the cathode GDL. Due to the hydrophobic property of the GDL, the liquid phase pressure in a cathode is larger than the gas phase pressure (capillary effect),[8] and a significant amount of liquid water is likely to be always maintained in the CAL, which makes Nafion ionomer fully hydrated. If Nafion ionomer is fully hydrated, the proton concentration is uniform in the CAL since the proton is the only ionic species in the electrolyte for charge transfer (the anion is immobile).[9] In contrast to a traditional alkaline fuel cell or a phosphoric acid fuel cell where the concentration variation of the electrolyte is important, the proton concentration in the CAL is not a variable in a PEMFC cathode model.[9] Therefore, this concentration was not explicitly included in this work. Similar to Springer et al.'s work,[1,7] the volume fractions of gas pores in both the GDL and the CAL were assumed not to change with the change of the operating current density, for simplicity. Due to this assumption, the transport of liquid water in the cathode was also not included in this work.

The procedures of making a membrane electrode assembly (MEA) in this work were similar to those described in the literature.[10] The Pt catalyst ink with 75 wt% catalyst and 25 wt% Nafion® ionomer (dry content) was prepared with an experimentally available 40.2 wt% Pt/Vulcan XC-72 catalyst (E-TEK Division, De Nora North America, NJ) and a perfluorosulfonic acid-copolymer (Alfa Asesar, MA). The ink was mixed properly for at least 8 hours. ELAT® GDLs (E-TEK Division, De Nora North America, NJ), which thickness was measured to be approximately 400 μm, were cut into 3.2×3.2 cm$^2$ pieces. The catalyst ink was sprayed onto the GDLs, and dried for ½ hour to evaporate any remaining solvent. This process was repeated until the target loading was achieved. The catalyzed GDLs, which served as both the anode and the cathode, were calculated to have a Pt loading of 0.5 mg/cm$^2$ and measured to have a CAL thickness of 15 μm. To make a MEA, two pieces of catalyzed GDLs were bonded to a pretreated Nafion® 112 membrane by hot pressing at 140 °C for two minutes under a pressure of 500 psig. The MEA was assembled into a test fuel cell with single channel serpentine flow field graphite end plates purchased from Fuel Cell Technologies.

## Cathode Model

With the assumption that isothermal, isobaric and equilibrium water vapor saturation conditions hold for a PEMFC air cathode, we developed in a previous work a steady state polarization model.[8] In the cathode GDL, the Stefan-Maxwell multi-component gas transport yields

$$\frac{\beta_1 + \beta_2 x}{(\beta_1 - x)(\beta_3 + \beta_2 x)} \frac{\partial x}{\partial z} = \frac{I}{4F\varphi_B^{1.5} D_{ON}^0 c_G / l_B} \tag{1}$$
$$\beta_1 = 1 - w, \beta_2 = D_{WN}^0 / D_{OW}^0 - 1, \beta_3 = 1 - w + w D_{WN}^0 / D_{OW}^0$$

where $x$ and $w$ are the steady state mole fractions of $O_2$ and water vapor in the air stream ($w$ is fixed due to the isothermal and equilibrium water vapor saturation conditions assumed), respectively, I is the steady state operating current density, $z$ is the spatial coordinate in the GDL normalized by its thickness $l_B$ (see Fig. 1), F is the Faraday's constant, $c_G$ is the total gas concentration, $\varphi_B$ is the porosity of the GDL, and $D_{ON}^0$, $D_{WN}^0$ and $D_{OW}^0$ are the binary diffusion coefficients of $O_2$-$N_2$, water vapor-$N_2$ and water vapor-$O_2$, respectively. If a constant value of $x$ at the GDL inlet is always maintained, equation 1 can be integrated analytically to yield

$$\frac{\beta_1\left(1+\beta_2\right)}{\beta_1\beta_2+\beta_3}\ln\left(\frac{\beta_1-x}{\beta_1-x_0}\right)+\frac{\beta_3-\beta_1}{\beta_1\beta_2+\beta_3}\ln\left(\frac{\beta_3+\beta_2x}{\beta_3+\beta_2x_0}\right)=-\frac{I}{4F\varphi_B^{1.5}D_{ON}^0c_G/l_B}z \qquad (2)$$

which has a form similar to equation 5 of Springer *et al.*'s work,[7] except that I has a negative sign here for the discharging process.[8]

In the cathode CAL, the Stefan-Maxwell multi-component gas transport yields

$$\frac{\beta_1+\beta_2x}{\left(\beta_1-x\right)\left(\beta_3+\beta_2x\right)}\frac{\partial^2 x}{\partial z^2}+\frac{\beta_4+2\beta_1\beta_2x+\beta_2^2x^2}{\left(\beta_1-x\right)^2\left(\beta_3+\beta_2x\right)^2}\left(\frac{\partial x}{\partial z}\right)^2=\frac{-j_O l_c}{\varphi_c^{1.5}D_{ON}^0c_G/l_c} \qquad (3)$$

$$\beta_4=\beta_1\beta_3-\beta_2\beta_1^2+\beta_1\beta_2\beta_3$$

where $z$ is the spatial coordinate in the CAL normalized by its thickness $l_c$, $\varphi_c$ is the porosity of the CAL, and $-j_O$ is the steady state consumption rate of $O_2$ gas

$$-j_O=3\left(1-\varphi_c\right)\frac{D_{eff}}{R_a^2}c_G x H\left(\sqrt{\frac{\frac{i_{ref}}{4Fc_{ref}}\exp\left(-\frac{\eta}{b}\right)}{\frac{D_{eff}}{R_a^2}}}\coth\left(\sqrt{\frac{\frac{i_{ref}}{4Fc_{ref}}\exp\left(-\frac{\eta}{b}\right)}{\frac{D_{eff}}{R_a^2}}}\right)-1\right) \qquad (4)$$

where $D_{eff}$ is the effective diffusion coefficient of $O_2$ in a flooded agglomerate particle, $R_a$ is the radius of that particle (In refs. 11 and 12, $R_a$ was measured to have an approximate value of 0.1 $\mu$m by using the scanning electron microscopy or the transmission electron microscopy technique.), H is the Henry's constant, $i_{ref}$ is the exchange current density of the $O_2$ reduction reaction per unit volume of the agglomerate particles at a reference liquid phase $O_2$ concentration $c_{ref}$ equal to $1.0\times10^{-6}$ mol/cm$^3$ (an equilibrium liquid phase $O_2$ concentration when the hydrated Nafion is exposed to $O_2$ gas with a pressure of around 1.0 atm), $b$ is the normal Tafel slope, and $\eta$ is the over-potential. Equation 4 is obtained by solving the steady state spherical diffusion inside an agglomerate particle and by assuming that the overall $O_2$ reduction reaction follows a four-electron mechanism:

$$O_2 + 4H^+ + 4e^- \rightarrow 2H_2O(l) \qquad (5)$$

Equation 2 can be used to find $x$ at the GDL/CAL interface to provide a boundary condition for equation 3 since

$$x\big|_{z=1,c} = x\big|_{z=1,B} \qquad (6)$$

Another boundary condition for equation 3 is

$$\frac{\partial x}{\partial z}\bigg|_{z=1,c} = 0 \qquad (7)$$

Equation 7 is obtained by assuming zero $O_2$ flux at the CAL/PEM interface.

A combination of the modified Ohm's law and the conservation of charge yields [8]

$$\frac{\partial^2 \eta}{\partial z^2} = \frac{l_c}{\kappa_{eff}} 4F j_O l_c - \frac{RT}{4F} \frac{\partial^2 \ln x}{\partial z^2} \qquad (8)$$

where $\kappa_{eff}$ is the effective ionic conductivity of the electrolyte, R is the universal gas constant, and T is the temperature in K. To obtain equation 8, an infinitely large electronic conductivity is assumed for the solid phase, and a hypothetical $O_2$ reference electrode placed right outside the surface of a flooded agglomerate particle is used to measure the electrolyte potential.

Equation 8 is subject to the following boundary conditions

$$\frac{\partial \eta}{\partial z}\bigg|_{z=0,c} = -\frac{RT}{4F} \frac{\partial \ln x}{\partial z}\bigg|_{z=0,c} \qquad (9)$$

and

$$\frac{\partial \eta}{\partial z}\bigg|_{z=1,c} = \frac{l_c}{\kappa_{eff}} I \tag{10}$$

The cathode potential in reference to a standard $H_2$ electrode is determined by the solid phase potential

$$\Phi_1 = (\eta + E)\big|_{z=1,c} \tag{11}$$

where E is the local equilibrium potential of the cathode and has a Nernst form

$$E = E_O^0 + \frac{RT}{4F}\ln(Px) \tag{12}$$

where $E_O^0$ is the standard potential of the cathode in reference to a standard $H_2$ electrode and P is the total gas pressure in atm.

It is noted that the numerical calculation of the steady state polarization data of a PEMFC air cathode is simplified to only one region, the CAL, since the solution of $x$ at the GDL/CAL interface is obtained analytically (see equation 2).

In this work, we are interested in estimating five parameters, $\varphi_B$, $\varphi_c$, $i_{ref}$, $D_{eff}/R_a^2$ and $\kappa_{eff}$, from the experimental polarization curves of a PEMFC air cathode by using the PEMFC cathode model described above.

## Nonlinear Parameter Estimation

Three least square methods are available for nonlinear parameter estimation: the steepest descent method, the Gauss-Newton method, and the Marquardt method.[13] The steepest descent method has the advantage of guaranteeing that the sum of the squared residuals $S^2$ will move toward its minimum without diverging but the disadvantage of slow convergence when $S^2$ approaches its minimum, while the Gauss-Newton method has the advantage of fast convergence when $S^2$ approaches its minimum but the disadvantage of diverging if the initial guesses of all the parameters are not very close to their final estimates. The Marquardt method is an interpolation technique between the Gauss-Newton and the steepest descent methods. It has the advantages of these two methods but none of their disadvantages. In general, the Marquardt method is associated with finding the parameter correction vector $\Delta\boldsymbol{\theta}$ [13]

$$\Delta\boldsymbol{\theta} = \left(\mathbf{J}^{\mathrm{T}}\mathbf{J} + \lambda\mathbf{I}\right)^{-1}\mathbf{J}^{\mathrm{T}}\left(\mathbf{Y}^* - \mathbf{Y}\right) \tag{13}$$

where $\mathbf{J}$ is a matrix of the partial derivatives of the dependent variable of a model with respect to estimation parameters evaluated at all the experimental data points, $\mathbf{Y}$ is the model prediction vector of the dependent variable, $\mathbf{Y}^*$ is the experimental observation vector of the dependent variable, $\lambda$ is the step size correction factor, $\mathbf{I}$ is the identity matrix, and the superscripts T and -1 are used to represent the transpose and inverse of a matrix, respectively. The sum of the squared residuals $S^2$ (un-weighted) is calculated by

$$S^2 = \left(\mathbf{Y}^* - \mathbf{Y}\right)^{\mathrm{T}}\left(\mathbf{Y}^* - \mathbf{Y}\right) \tag{14}$$

An algorithm of the Marquardt method consists of the following steps: (i) assume initial guesses for the parameter vector $\boldsymbol{\theta}$; (ii) assign a large value, *i.e.*, 1000, to $\lambda$ to assure that initial parameter corrections will move toward the lowered sums of the squared residuals; (iii) evaluate $\mathbf{J}$; (iv) use equation 13 to obtain $\Delta\boldsymbol{\theta}$; (v) calculate the updated $\boldsymbol{\theta}$ by

$$\boldsymbol{\theta}^{(m+1)} = \boldsymbol{\theta}^{(m)} + \Delta\boldsymbol{\theta}^{(m)} \tag{15}$$

where the superscript m represents the correction number; (vi) calculate $S^2$, and reduce the value of $\lambda$ if $S^2$ is decreased or increase the value of $\lambda$ if $S^2$ is increased; (vii) repeat steps (iii)-(vi) until either $S^2$ does not change appreciably or $\Delta\theta$ becomes very small or both are satisfied.[13]

For a model involving differential equations, the accurate calculation of $\mathbf{J}$ is very important for avoiding diverging in the parameter estimation process. There are two ways to calculate $\mathbf{J}$: the finite difference approach and the sensitivity approach.[14] A simple way to calculate $\mathbf{J}_{ij}$ at a data point i by using the finite difference approach is the one-sided approximation:

$$\mathbf{J}_{ij} = \frac{\mathbf{Y}_i\left(...,\boldsymbol{\theta}_j+\Delta\boldsymbol{\theta}_j,...\right) - \mathbf{Y}_i\left(...,\boldsymbol{\theta}_j,...\right)}{\Delta\boldsymbol{\theta}_j} \qquad (16)$$

The main advantage of this approach is its convenience in coding. However, large error is sometimes generated. Two sources of error contribute to the inaccuracy of finding $\mathbf{J}_{ij}$ from equation 16: the rounding error arising when two closely spaced values of $\mathbf{Y}_i$ are subtracted from each other and the truncation error due to the inexact nature of equation 16, which is accurate only when $\Delta\boldsymbol{\theta}_j \rightarrow 0$.[14] While the truncation error decreases with the decrease of $\Delta\boldsymbol{\theta}_j$, the rounding error increases. A central finite difference approximation may be helpful to reduce the truncation error. Unfortunately, additional numerical solutions of model equations are required compared to the one-sided approximation while the rounding error may be still significant. To eliminate the rounding error completely in the calculation of $\mathbf{J}$, the sensitivity approach is very useful. In contrast to the finite difference approach, the sensitivity approach calculates directly the derivative of a state variable with respect to a parameter, which is called the sensitivity coefficient.[14] To demonstrate, let us consider a case that the volume fraction of gas pores in the CAL, $\varphi_c$, is to be estimated alone by using the model described in the previous session. By taking the partial derivatives with respect to $\varphi_c$ on both sides of equation 3, we obtain

$$\frac{\beta_1+\beta_2 x}{(\beta_1-x)(\beta_3+\beta_2 x)}\frac{\partial^2 S_{x,\varphi_c}}{\partial z^2}+\frac{\beta_4+2\beta_1\beta_2 x+\beta_2^2 x^2}{(\beta_1-x)^2(\beta_3+\beta_2 x)^2}\left(\frac{\partial^2 x}{\partial z^2}S_{x,\varphi_c}+2\left(\frac{\partial x}{\partial z}\right)\frac{\partial S_{x,\varphi_c}}{\partial z}\right)$$

$$+2\frac{\beta_5+\beta_2^3 x^3+3\beta_1\beta_2^2 x^2+\left(\beta_1\beta_2\beta_3-\beta_1^2\beta_2^2+\beta_1\beta_2^2\beta_3+2\beta_2\beta_4\right)x}{(\beta_1-x)^3(\beta_3+\beta_2 x)^3}\left(\frac{\partial x}{\partial z}\right)^2 S_{x,\varphi_c} \qquad (17)$$

$$=\frac{-j_O l_c}{\varphi_c^{1.5}D_{ON}^0 c_G/l_c}\left(\frac{S_{x,\varphi_c}}{x}-\frac{1.5}{\varphi_c}-\frac{1}{1-\varphi_c}-\frac{\sqrt{k}\coth\left(\sqrt{k}\right)+k-k\coth\left(\sqrt{k}\right)^2}{2b\left[\sqrt{k}\coth\left(\sqrt{k}\right)-1\right]}S_{\eta,\varphi_c}\right)$$

where

$$\beta_5=\beta_1^2\beta_2\beta_3+\beta_3\beta_4-\beta_1\beta_2\beta_4,$$

$$k=\frac{i_{ref}/\left(4Fc_{ref}\right)}{D_{eff}/R_a^2}\exp\left(-\frac{\eta}{b}\right), \qquad (18)$$

$$S_{\eta,\varphi_c}=\frac{\partial\eta}{\partial\varphi_c} \text{ and } S_{x,\varphi_c}=\frac{\partial x}{\partial\varphi_c}$$

By substituting $z=1$ into equation 2 and taking the partial derivatives with respect to $\varphi_c$ on both sides, we obtain a boundary condition for equation 17:

$$\left[\frac{\beta_2\left(\beta_3-\beta_1\right)}{\left(\beta_3+\beta_2 x\big|_{z=1,B}\right)}-\frac{\beta_1\left(1+\beta_2\right)}{\left(\beta_1-x\big|_{z=1,B}\right)}\right]\frac{S_{x,\varphi_c}\big|_{z=0,c}}{\beta_1\beta_2+\beta_3}=-\frac{I}{4F\varphi_B^{1.5}D_{ON}^0 c_G/l_B} \qquad (19)$$

By taking the partial derivatives with respect to $\varphi_c$ on both sides of equation 7, we obtain another boundary condition for equation 17:

$$\frac{\partial S_{x,\varphi_c}}{\partial z}\bigg|_{z=1,c}=0 \qquad (20)$$

Similarly, by taking the partial derivatives with respect to $\varphi_c$ on both sides of equations 8-10, we obtain

$$\frac{\partial^2 S_{\eta,\varphi_c}}{\partial z^2} + \frac{RT}{4F}\frac{\partial^2\left(\dfrac{S_{x,\varphi_c}}{x}\right)}{\partial z^2} = \frac{l_c}{\kappa_{eff}}4Fj_O l_c$$

$$\times\left(\frac{S_{x,\varphi_c}}{x} - \frac{1}{1-\varphi_c} - \frac{\sqrt{k}\coth\left(\sqrt{k}\right)+k-k\coth\left(\sqrt{k}\right)^2}{2b\left[\sqrt{k}\coth\left(\sqrt{k}\right)-1\right]}S_{\eta,\varphi_c}\right) \tag{21}$$

$$\left.\frac{\partial S_{\eta,\varphi_c}}{\partial z}\right|_{z=0,c} = -\frac{RT}{4F}\left.\frac{\partial\left(\dfrac{S_{x,\varphi_c}}{x}\right)}{\partial z}\right|_{z=0,c} \tag{22}$$

and

$$\left.\frac{\partial S_{\eta,\varphi_c}}{\partial z}\right|_{z=1,c} = 0 \tag{23}$$

The sensitivity coefficients $S_{x,\varphi_c}$ and $S_{\eta,\varphi_c}$ can be solved numerically from equations 17 and 19-23, which are called the sensitivity equations,[14] if the profiles of $x$ and $\eta$ are known. After taking the partial derivatives with respect to $\varphi_c$ on both sides of equation 11, we can calculate, $\mathbf{J}_i$, the partial derivative of the dependent variable $\Phi_1$ with respect to $\varphi_c$ at a current density data point i

$$\mathbf{J}_i = \left(\frac{\partial\Phi_1}{\partial\varphi_c}\right)_i = \left.S_{\eta,\varphi_c}\right|_{z=1,c} + \frac{RT}{4F}\left.\left(\frac{S_{x,\varphi_c}}{x}\right)\right|_{z=1,c} \tag{24}$$

If several parameters are to be estimated together, in a similar manner, we can obtain some corresponding sensitivity equations and calculate $\mathbf{J}_{ij}$, the partial derivative of the dependent variable $\Phi_1$ with respect to parameter $\theta_j$ at a current density data point i:

$$\mathbf{J}_{ij} = \left(\frac{\partial\Phi_1}{\partial\theta_j}\right)_i = \left.S_{\eta,\theta_j}\right|_{z=1,c} + \frac{RT}{4F}\left.\left(\frac{S_{x,\theta_j}}{x}\right)\right|_{z=1,c} \tag{25}$$

The main advantage of the sensitivity approach is its accuracy in finding **J** without possibly demanding more computer time, even if it is less friendly for coding compared to the finite difference approach.

In this work, the Marquardt method was combined with the sensitivity approach for the estimation of parameters of interest from the experimental steady state polarization data of a PEMFC air cathode. After scrutinizing the model equations described in the previous session, we find that $\varphi_B$, $\varphi_c$, $i_{ref}$, $D_{eff}/R_a^2$ and $\kappa_{eff}$ are very important parameters and the values of them should be obtained before the accurate prediction of a cathode performance is possible. Among them, $\varphi_B$, $\varphi_c$, $i_{ref}$ and $\kappa_{eff}$ are the physical meaningful parameters, and the reciprocal of $D_{eff}/R_a^2$ can be interpreted as the time constant for $O_2$ diffusion inside a flooded agglomerate particle.

The normal Tafel slope $b$ is a kinetics parameter, which value was measured and reported in the literature.[15-19] This parameter was not included in our estimation. The thicknesses of the GDL and the CAL were measured on a gas diffusion electrode. They were also not included in our estimation.

From the statistics point of view, it is more desirable to obtain a confidence interval of a parameter rather than to simply obtain its point estimate. In this work, the 95% confidence interval of a parameter $\theta_j$ is constructed by [13]

$$\theta_j^* - t_{(1-0.05/2)} S_E \sqrt{\mathbf{a}_{jj}} \leq \theta_j < \theta_j^* + t_{(1-0.05/2)} S_E \sqrt{\mathbf{a}_{jj}} \qquad (26)$$

where $\theta_j^*$ represents the point estimate of parameter $\theta_j$, $t_{(1-0.05/2)}$ is a value of the student's $t$ distribution with (n-m) degrees of freedom where n and m are the numbers of experimental data points and estimation parameters, respectively, $\mathbf{a}_{jj}$ is a diagonal element of the matrix $(\mathbf{J}^T\mathbf{J})^{-1}$, and $S_E$ is an unbiased estimate of the variance and can be calculated by

$$S_E^2 = \frac{\sum_{i=1}^{n}\left[(\Phi_1)_i - (\Phi_1^*)_i\right]^2}{n-m} \tag{27}$$

where $\Phi_1^*$ is the experimental cathode potential. For a nonlinear model, due to correlations between parameter pairs, the calculated confidence intervals are not as rigorous as those for a linear model, and a joint confidence region of all the estimation parameters is more useful for identifying their true region. The 95% joint confidence region of estimation parameters can be obtained by [13]

$$\frac{\left(\boldsymbol{\theta}^* - \boldsymbol{\theta}\right)^T\left(\mathbf{J}^T\mathbf{J}\right)\left(\boldsymbol{\theta}^* - \boldsymbol{\theta}\right)}{mS_E^2} \le F_{(1-0.05)}\left(m, n-m\right) \tag{28}$$

where $F_{(1-0.05)}(m,\ n\text{-}m)$ is a value of the $F$ distribution with m and (n-m) degrees of freedom.

## Numerical Method

A three-point finite difference method was used to approximate each derivative variable in a differential equation, and a general nonlinear equation solver in Fortran called GNES was used to carry out all the numerical calculations. An important feature of this solver is its convenience in coding and debugging. Normally, only the model equations are required from a user. The Jacobian matrix for numerical calculation is not required, since the solver can generate it internally by using a forward finite difference approximation method. To improve computation efficiency, however, a user may provide a banded Jacobian matrix to the solver.

To find the parameter correction vector $\Delta\boldsymbol{\theta}$ by using equation 13, one needs to calculate the model prediction vector $\mathbf{Y}$ as well as the matrix $\mathbf{J}$. Therefore, the numerical solutions of $\Phi_1, \partial\Phi_1/\partial\varphi_B, \partial\Phi_1/\partial\varphi_c, \partial\Phi_1/\partial i_{ref}, \partial\Phi_1/\partial\left(D_{eff}/\mathrm{R}_a{}^2\right)$ and $\partial\Phi_1/\partial\kappa_{eff}$ at each current density data point were required. We elected not to couple five sets of sensitivity equations such as equations 17 and 19-23 to the original model equations and solve them simultaneously in our calculations. The decoupling of model equations from sensitivity equations saves computer time due to the following concerns: (i) The computer time required for performing the LU decomposition on six matrices of the same size, *i.e.*, n×n, is less than that required for performing the decomposition on a single matrix of a sixfold size, *i.e.*, 6n×6n (the LU decomposition method is used by GNES in its numerical calculation); (2) The coupling of five sets of sensitivity equations, which are linear with respect to all the sensitivity coefficients and do not require iterations for their numerical solutions, to the model equations, which are nonlinear with respect to their state variables such as $x$ and $\eta$ and require iterations for their numerical solutions, will inevitably force all the sensitivity equations to undergo the same number of iterations before all the converged solutions are obtained. An efficient numerical algorithm is very important for a nonlinear parameter estimation problem with a sophisticated differential equation model such as the model considered in this work, since a great number of numerical calculations are usually necessary before the final parameter estimates are determined. After providing a banded Jacobian matrix to the solver and calculating the model equations (to be solved first) and each set of sensitivity equations separately, only 10 min was taken by a personal computer with an 866 MHz CPU to obtain 10 parameter correction vectors. (84 experimental data point were considered, and 100 node points were used to discretize the spatial coordinate $z$.)

## Experimental

The test fuel cell was operated on a 120 A fuel cell test station (Fuel Cell Technologies). The temperatures of the test cell and the cathode gas humidifier were set to be 70 °C, while the temperature of the anode gas humidifier was set to be 10 °C more in order to avoid the partial dehydration of the PEM on the anode side. The test fuel cell was first operated at 0.6V under the ambient gas pressure for at least 8 hours with a 250 cm$^3$/min $O_2$ flow rate on the cathode side and a 180 cm$^3$/min $H_2$ flow rate on the anode side. Then the cathode gas feeding was switched to air with a flow rate of 720 cm$^3$/min. The flow rate of $H_2$ was increased to be 640 cm$^3$/min. High flow rates on both the cathode and the anode were employed in this work in order to maintain a constant mole fraction of $O_2$ at the cathode GDL inlet as well as to support the largest current attainable on an air/$H_2$ PEMFC during the steady state polarization curve measurements. The anode gas pressure was set to be 1.3 atm, a value that makes the partial pressure of $H_2$ in the anode gas pores equal to 1.0 atm, while three different values, 1.3, 2.3 and 3.3 atms, were used for the cathode gas pressures. After a new cathode gas pressure was set, the cell was first operated at 0.6 V for at least 30 min, and then a steady state polarization curve was measured. To measure a polarization curve of a PEMFC, the cell potential was swept from 1.0 to 0.1 and to 1.0 V with a step size of 25 mV and a delay time of 15 s. To obtain a polarization curve of the air cathode, the voltage drop across the PEM was used to correct the polarization curve of a full cell. Due to the fact that the PEM resistance is unlikely to be a strong function of the operating current density if a thin PEM is used and a good gas humidification of the anode is always guaranteed, we assumed the existence of a constant value of the PEM resistance in this work during each polarization curve measurement and used a simple Ohm's law to calculate the voltage drop across the PEM at each current density data point. The PEM resistance was measured at 10 KHz with a Hewlett Packard/Agilent 4263B LCR meter at the open circuit conditions immediately after each polarization curve was measured. In this work, the same value of 78 m$\Omega$-cm$^2$ was obtained for the PEM resistance in all the measurements.

## Results and Discussion

In our model, the values of some parameters such as $D_{ON}^0$, $D_{OW}^0$, $D_{NW}^0$, $l_B$, $l_c$, $b$, H and $E_O^0$ can be obtained accurately from either direct measurements or the literature.[15-20] They are presented in Table I. The remaining five parameters, $\varphi_B$, $\varphi_c$, $i_{ref}$, $D_{eff}/R_a^2$, and $\kappa_{eff}$ have to be estimated from the experimental polarization curves. Springer *et al.*[1] suggested that the simultaneous fit of several sets of experimental data measured under different operating conditions provides one with more effective diagnostics than it is possible from a fit of only one set of experimental data at a time. In this work, our model was used to fit three experimental polarization curves of an air cathode simultaneously. To demonstrate the goodness of the simultaneous fit, the model was also used to fit each experimental curve independently, for comparison purposes. The 95% confidence intervals of all the five parameters obtained from the simultaneous fit are presented in Table II. The polarization curve predictions of the simultaneous fit are compared with three experimental curves in Figs. 2-1 and 2-2. In general, a good match of model predictions with experimental curves can be observed from these two figures. Therefore, the simultaneous fit was performed effectively.

One may want to know whether or not there is further improvement of a fit if only one experimental curve is considered at a time for the parameter estimation. The 95% confidence intervals of all the five parameters obtained from three independent fits are also presented in Table II. The polarization curve predictions of these independent fits are compared with experimental curves in Fig. 3. Even if Table II shows that each independent fit leads to a smaller $S_E$ compared to the simultaneous fit, it is hard for one to simply conclude that Fig. 3 displays much better fit than Fig. 2-1.

One may notice from the results of three independent fits presented in Table II that with the decrease of the cathode gas pressure, the value of $\kappa_{eff}$ decreases, while the values of $i_{ref}$ and $D_{eff}/R_a^2$ increase. An exclusive explanation for all these phenomena is very difficult to find. One may attribute the decrease of $\kappa_{eff}$ to the expansion effect of the CAL thickness with the decrease of gas pressure. Unfortunately, the increases of $D_{eff}/R_a^2$ and $i_{ref}$ can not be answered properly by this explanation. Alternately, one may attribute the decrease of $\kappa_{eff}$ and the increase of $D_{eff}/R_a^2$ to the partial Nafion ionomer dehydration in the CAL with the decrease of gas pressure (Due to insufficient water content, very small gas pores may be left open in an agglomerate particle under a low gas pressure to facilitate $O_2$ diffusion to the catalyst sites.). However, the increase of $i_{ref}$ with the decrease of gas pressure cannot be explained. As noticed from Figs. 2-1 and 3, our model predictions match experimental curves not very well in the medium current density range. The understanding of this phenomenon is probably useful to explain the changes of $\kappa_{eff}$, $i_{ref}$ and $D_{eff}/R_a^2$ with the change of gas pressure. We recall that the values of $\varphi_B$ and $\varphi_c$ were assumed to be independent of the operating current density in this work. Rigorously speaking, it is not true. A small operating current density is expected to incur a small liquid water flux out of the cathode GDL and consequently cause a small number of gas pores to be flooded. A large operating current density is expected to incur a large liquid water flux out of the GDL and consequently cause a great number of gas pores to be flooded. Therefore, the values of $\varphi_B$ and $\varphi_c$ in the medium current density range are expected to be larger than those in the high current density range. Even if the extracted values of $\varphi_B$ and $\varphi_c$ presented in Table II are not noticed to vary much with the change of gas pressure, the possibility that these values change with the operating current density is not excluded. A proper modeling of the transport of liquid water in both the GDL and the CAL in a manner similar to that introduced in ref. 5, where the Darcy's law was used for this purpose, is expected to take into account the changes of $\varphi_B$ and $\varphi_c$ with the change of current density and improve our polarization curve predictions. In this work, all the experimental polarization curves of a PEMFC were measured by sweeping the cell potential in both the forward and backward directions, and an effort to discriminate part of experimental data obtained from a particular direction over the other was not attempted. Because of this, there were appreciable differences between the experimental data measured in two potential sweep directions in the medium current density range. These differences could be explained by the hysteresis behavior of the performance of a PEMFC cathode associated with liquid water inhibition and drainage in the GDL.[21-23] This hysteresis behavior, which was particularly significant for a low-pressure cathode (see Figs. 2-1 and 3), introduced appreciable noise to our experimental data.

Once may also notice from Table II that the confidence interval of $D_{\text{eff}}/R_a^2$ is much larger than that of any of the other four parameters. This indicates uncertainty in the determination of $D_{\text{eff}}/R_a^2$. A large confidence interval of a parameter was also obtained by Evans and White.[24] They explained that an unacceptably large confidence interval of a parameter was related to parameter correlations in a nonlinear model. To verify this explanation, we fixed all the other four parameters and estimated the parameter $D_{eff}/R_a{}^2$ from a simultaneous fit of three experimental curves. Since only one parameter was left for estimation, parameter correlations were removed. As expected, in the absence of parameter correlations, a much smaller confidence interval of $D_{eff}/R_a{}^2$ was obtained: $2.792 \times 10^3 \leq D_{eff}/R_a{}^2 < 3.312 \times 10^3$ s$^{-1}$.

The degree of correlation between any two parameters in our nonlinear model can be appreciated by looking at the correlation coefficient matrix $\mathbf{R}$ obtained from $(\mathbf{J}^T\mathbf{J})^{-1}$ (see ref. 13) after the simultaneous fit:

$$\mathbf{R} = \begin{bmatrix} 1.000 & 0.5176 & 0.3113 & -0.05743 & -0.9070 \\ 0.5176 & 1.000 & 0.3357 & -0.6786 & -0.4223 \\ 0.3113 & 0.3357 & 1.000 & -0.5072 & -0.1819 \\ -0.05743 & -0.6786 & -0.5072 & 1.000 & -0.2339 \\ -0.9070 & -0.4223 & -0.1819 & -0.2339 & 1.000 \end{bmatrix} \qquad (29)$$

where for either subscript of the element $\mathbf{R}_{ij}$, "1" represents $\varphi_B$, "2" represents $\varphi_c$, "3" represents $i_{ref}$, "4" represents $D_{eff}/R_a{}^2$, and "5" represents $\kappa_{eff}$.

As explained in ref. 13, the higher the correlation between two parameters, the closer the absolute value of $\mathbf{R}_{ij}$ is to 1.0. One can observe from equation 29 that the values of all the diagonal elements of $\mathbf{R}$ are equal to 1.0. This indicates that each parameter is highly correlated with itself. One can also observe from equation 29 that the highest correlation between two different parameters occurs to the $\varphi_B$-$\kappa_{eff}$ pair, and the lowest correlation between two different parameters occurs to the $\varphi_B$-$D_{eff}/R_a{}^2$ pair. The correlations between the $\varphi_c$-$D_{eff}/R_a{}^2$ pair, the $i_{ref}$ -$D_{eff}/R_a{}^2$ pair and the $\varphi_B$-$\varphi_c$ pair are also high. Ref. 13 explains that a positive correlation coefficient between two parameters implies that the errors causing the estimate of one parameter to be high also cause the other to be high, and a negative correlation coefficient implies that the errors causing the estimate of one parameter to be high cause the other to be low. Since the $\varphi_B$-$\kappa_{eff}$ pair has a very negative correlation coefficient, it is not difficult for one to conclude that if $\kappa_{eff}$ was underestimated in this work, an overestimation of $\varphi_B$ resulted.

We know from ref. 13 that for a linear model, all the estimation parameters are uncorrelated, the axes of the confidence ellipsoid is parallel to the coordinates of the parameter space, and the individual parameter confidence intervals hold for each parameter independently; whereas for a nonlinear model, the parameters are correlated, the axes of the confidence ellipsoids are at an angle to the parameter space, and the individual parameter confidence limits do not represent the true interval within which a parameter may lie. Therefore, the confidence intervals presented in Table II are not rigorously valid, and a joint confidence region for all the parameters is more appropriate. In this work, the 95% joint confidence region for all the five parameters estimated from the simultaneous fit is obtained by using equations 30-31:

$$(\Delta\boldsymbol{\theta})^{\mathrm{T}} \begin{bmatrix} 3.768\times10^4 & 7.056\times10^3 & 7.095\times10^4 & 1.559\times10^{-2} & 3.739\times10^4 \\ 7.056\times10^3 & 2.033\times10^3 & 1.995\times10^4 & 4.036\times10^{-3} & 8.298\times10^3 \\ 7.095\times10^4 & 1.995\times10^4 & 3.307\times10^5 & 4.373\times10^{-2} & 8.604\times10^4 \\ 1.559\times10^{-2} & 4.036\times10^{-3} & 4.373\times10^{-2} & 8.548\times10^{-9} & 1.769\times10^{-2} \\ 3.739\times10^4 & 8.298\times10^3 & 8.604\times10^4 & 1.769\times10^{-2} & 4.017\times10^3 \end{bmatrix} (\Delta\boldsymbol{\theta}) \leq 1.729\times10^{-3} \quad (30)$$

where

$$\Delta\boldsymbol{\theta} = \begin{bmatrix} \varphi_B - 0.1991 \\ \varphi_c - 3.933 \times 10^{-2} \\ i_{ref} - 7.198 \times 10^{-4} \\ D_{eff}/R_a^2 - 3.052 \times 10^3 \\ \kappa_{eff} - 9.947 \times 10^{-3} \end{bmatrix} \tag{31}$$

The disadvantage of using equations 30-31 is the lack of straightforwardness in identifying the confidence region where all the parameters lie. One may fix the values of some parameters, and determine the confidence region for the remaining parameters. For instance, if the values of $\varphi_B$, $\varphi_c$, $i_{ref}$ and $\kappa_{eff}$ in equations 30-31 are fixed to their respective point estimates obtained from the simultaneous fit, one can obtain the confidence region for $D_{eff}/R_a^2$:

$$2.603 \times 10^3 \leq D_{eff}/R_a^2 < 3.502 \times 10^3 \text{ s}^{-1} \tag{32}$$

To appreciate the goodness of the polarization curve predictions by using a parameter value defined by a joint confidence region rather than by a confidence interval, a comparison of several simulated polarization curves of the medium-pressure air cathode (P=2.3 atm) is shown in Figs. 4-1 and 4-2. While the values of all the other four parameters in the polarization curve simulations were fixed to their respective point estimates obtained from the simultaneous fit, the values of $D_{eff}/R_a^2$ were assigned by the upper and lower limits defined by its 95% confidence interval as well as those defined by equation 32. One can notice from these two figures that the limits defined by the joint confidence region (equation 32) leads to less degree of uncertainty in the model predictions than those defined by the confidence interval of $D_{eff}/R_a^2$.

If PEMFCs are widely used to power the electric vehicles in the future, their cathodes are very likely going to be operated with low-pressure air due to the energy cost of gas pressurizing. Therefore, a proper understanding of mass transport limitations of a low-pressure PEMFC cathode is very important. The distributions of the mole fraction of $O_2$ across the CAL of the low-pressure air cathode (P=1.3 atm) operated at different current densities are presented in Fig. 5. The point estimates obtained from the simultaneous fit were used by their corresponding parameters for the calculation of all the $x$ distributions. In general, the value of $x$ decreases in the direction toward the PEM. With the increase of the operating current density, the value of $x$ at the GDL/CAL interface also decreases due to the gas phase transport loss of $O_2$ in the GDL.[8] When the current density increases to a value as high as 1.5 A/cm$^2$, except for a small region close to the GDL/CAL interface, all the other CAL region has a negligible $O_2$ content. As noticed in Fig. 2-1, the value of 1.5 A/cm$^2$ is close to the limiting current of the low-pressure air cathode (P=1.3 atm). Therefore, the gas phase transport limitation across the GDL is responsible for a limiting current measured on an air cathode. Similar conclusion was also drawn in the literature.[1,4]

Another way to understand mass transport limitations in the low-pressure air cathode (P=1.3 atm) is to look at the $O_2$ reduction current distributions in the CAL. The dimensionless $4Fj_0l_c/I$ *vs. z* plots are presented in Fig. 6 with the change of the operating current density. When the current density is very low, *i.e.*, -I=0.05 A/cm$^2$, an almost uniform distribution of $O_2$ reduction current exists. At this current density, the cathode performance is dominated only by slow Tafel kinetics.[3] When the current density becomes higher, *i.e.*, -I=0.5 A/cm$^2$, a non-uniform distribution of $O_2$ reduction current in the CAL is observed, and the reaction at the CAL/PEM interface is favored. At this current density, the cathode performance is very likely controlled by both processes: slow ionic conduction and slow Tafel kinetics (to be justified later).[3] When the current density becomes even higher, *i.e.*, -I=1.2 A/cm$^2$, high $O_2$ reduction current is seen not only in a region close to the CAL/PEM interface but also in a region close to the GDL/CAL interface. At this current density, the cathode performance is likely controlled jointly by slow gas phase mass transport and slow ionic conduction (to be justified later).[3] When the current density is as high as 1.5 A/cm$^2$, $O_2$ reduction reaction occurs predominately at the GDL/CAL interface. At this current density, $O_2$ gas is depleted in most of the CAL except for a small region close to the GDL/CAL interface (Fig. 5), and the cathode performance is mainly influenced by the gas phase transport limitation across the GDL.[1]

To gain further understanding as to how the performance of a cathode is dominated by one or more slow processes with the change of current density, it is helpful to look at Fig. 7, where the simulated steady state polarization curve of a cathode fed with high-pressure air (P=5.1 atm) is compared to the simulated curves of three cathodes fed with low-pressure $O_2$ (P=1.3 atm). Two different values of gas pressure are chosen for the air cathode and the $O_2$ cathodes in the simulations so that the partial pressure of $O_2$ at the GDL inlet is the same (1 atm) and all the polarization curves agree in the low current density region where the sluggish Tafel kinetics is the only limiting process. Among the three $O_2$ cathodes, an infinitely large value of $\kappa_{eff}$ was assumed for one $O_2$ cathode, and the infinitely large values of both $\kappa_{eff}$ and $D_{eff}/R_a{}^2$ were assumed for another $O_2$ cathode. For the latter cathode, due to the disappearance of ionic conduction and liquid phase $O_2$ diffusion limitations, the cathode behaves like a planar electrode and a normal Tafel slope is always presented. For the former cathode, the cathode behaves like a thin-film diffusion electrode and the possible change of Tafel slope due to slow liquid phase $O_2$ diffusion is reflected. One may notice by comparing the polarization curves of three $O_2$ cathodes in Fig. 7 that for the $O_2$ cathode with all the parameter values obtained from the simultaneous fit in this work, the change of Tafel slope is mainly due to a limitation by slow ionic conduction, and the limitation by $O_2$ diffusion in an agglomerate particle seems to be insignificant until the current density is very high, *i.e.*, -I=10 A/cm$^2$. For the air cathode with all the parameter values obtained from the simultaneous fit in this work, the change of Tafel slope due to gas phase transport loss of $O_2$ is observed when the operating current density is not very small. It is also possible that the agglomerate particle diffusion of $O_2$ also limits the air cathode performance when the current density approaches the limiting current since the $O_2$ reduction reaction is limited to a very small region close to the GDL/CAL interface at this current density (see the curve with −I=1.5A/cm$^2$ in Fig. 6).

The optimization of a PEMFC is usually associated with overcoming one or more mass transport limitations. In this study, the influences of changing the values of some parameters on the cathode performance are briefly studied and presented in Fig. 8, where the point estimates of all the five parameters obtained from the simultaneous fit were used for the base case simulation, and only one parameter value was allowed to change from the base case for the simulation of any other curve. One can observe from this figure that any increase of $\varphi_B$, $\varphi_c$, $i_{ref}$, $\kappa_{eff}$ and $D_{eff}/R_a^2$ leads to an improvement of the cathode performance. Among them, the increase of $\varphi_B$ influences the limiting current value most effectively. One may ask whether or not a significant improvement of the performance of an air cathode is possible by using a GDL with a larger volume fraction of gas pores and a smaller thickness, since both of them lead to the decrease of gas phase transport loss of $O_2$. In one experiment, we tested a specially designed PEMFC by using a very porous, approximately 200 μm thick GDL (many large open pores were observed on the GDL against the light) to make the air cathode, and noticed that the performance of this cell was even worse than that observed on a cell with the use of a regular GDL to make the cathode. However, one should not simply conclude from this experiment that the decrease of the GDL thickness or the increase of the volume fraction of gas pores of the GDL does not lead to an improvement of the cathode performance. The presence of many large open pores in the GDL could be very harmful to the cathode, since large pores were likely to lead to the quick loss of liquid water in the CAL and consequently lead to the decrease of the electrolyte conductivity. We would like to believe that it is very important to maintain a sufficient amount of liquid water in the CAL to make Nafion ionomer fully hydrated. If one is able to make a thinner GDL without introducing many big open pores, a better performance of a cathode with such GDL should be expected. One can also observe from Fig. 8 that except for the current density range close to the limiting current value, the increase of $i_{ref}$ improves the cathode performance more significantly than the increase of any other parameter. This is because an increase of $i_{ref}$ is predicted by our model to cause the vertical translational movement of an entire polarization curve to a place at higher potentials.[8] The translational distance $\Delta\Phi_1$ due to the increase of $i_{ref}$, $\Delta i_{ref}$, can be determined by [8]

$$\Delta\Phi_1 = b \ln\left(1 + \frac{\Delta i_{ref}}{i_{ref}}\right) \tag{31}$$

Even if it seems that one can increase the value of $i_{ref}$ by increasing the weight percentage of the catalyst Pt in the Pt/C composites, it is tricky to realize this in practice, since with the increase of this weight percentage, the particle size of Pt tends to grow and the specific surface area of Pt tends to decrease.[25] If the value of $i_{ref}$ is proportional to the surface area of Pt per unit volume of the CAL, an increase of the weight percentage of Pt will not always guarantee the increase of $i_{ref}$. One can also observe from Fig. 8 that due to the increase of $\kappa_{eff}$, the cathode performance is improved very effectively in a wide range of the operating current density, whereas the improvement of the cathode performance due to the increase of either $D_{eff}/R_a^2$ or $\varphi_c$ is effective only in the high current density range. In our previous study of the $\kappa_{eff}$ profile of an air cathode,[26] we concluded that there was an optimal amount of Nafion ionomer loading in the CAL of a cathode (ELAT® electrode). Either too much or too small Nafion loading did not lead to a good performance of a cathode. Besides, a nonlinear ionic conductivity distribution in the cathode CAL was noticed. The existence of a nonlinear ionic conductivity distribution on an ELAT® electrode is understandable since Nafion ionomer was applied to the CAL by spraying and a gradient of Nafion ionomer loading was created in the CAL. Even if the technique used in this work to make a cathode is different from our previous work and a uniform ionic conductivity distribution in the cathode CAL is expected here, we would like to believe that an optimal amount of Nafion ionomer loading in a PEMFC cathode CAL will always be true. The cathode performance improvement with the increase of $D_{eff}/R_a^2$ can be explained by the decrease of the time constant for $O_2$ diffusion inside a flooded agglomerate particle. The possibility of observing the change of Tafel slope from a normal value to a double value associated with liquid phase $O_2$ diffusion process on a polarization curve of a PEMFC cathode was discussed extensively in the literature.[3,9] Interestingly, the change of Tafel slope was also observed in the kinetics studies of the catalyst Pt on a rotating disc electrode:[15-19] at high potentials a single Tafel slope is exhibited, and at low potentials a double Tafel slope is exhibited. The change of Tafel slope observed in the kinetics studies was explained by the change of $O_2$ reduction mechanism from a four-electron path to a two-electron path.[15-16]

To demonstrate how effectively our numerical algorithm is improved by calculating the model equations and each set of sensitivity equations separately and by providing a banded Jacobian matrix, the computer time required to solve our nonlinear model equations with the change of their Jacobian matrix property is summarized in Table III. Since there are only two equations in our model for each spatial node point, the calculation of 200 equations indicates the use of 100 node points to discretize the spatial coordinate $z$. By solving 200 equations six times (only one data point is considered), we want to simulate the total computer time required for solving the model equations and each set of sensitivity equations separately. By solving 1200 equations once, we want to simulate the computer time necessary for solving the coupled model and sensitivity equations together. Table III shows that the numerical efficiency associated with the separate calculation of equations is improved by only 20% if a sparse Jacobian matrix exists and it is provided. For the case that there exists a sparse Jacobian matrix but it is not provided, the numerical efficiency is improved by 70%. For the case that there exists a dense Jacobian matrix and it is not provided, the separate calculation improves the numerical efficiency by 83%. Since an improvement of numerical efficiency associated with the separate calculation is always true, this method should be recommended in a nonlinear parameter estimation problem involving the numerical solution of differential equations.

# Conclusions

The simultaneous fit of three experimental curves was performed successfully by using a nonlinear parameter estimation method and an optimized numerical algorithm. The 95% joint confidence region obtained for the five parameters of interest are found to be more appropriate for the determination of their true parameter values rather than the 95% confidence intervals.

# Acknowledgements

The authors are grateful for the financial support of the project for Hybrid Advanced Power Sources by the National Reconnaissance Office (NRO) under Contract No. NRO-000-01-C-4368.

# List of Symbols

$b$      Normal Tafel slope, V

$c_G$      Total gas concentration, $mol/cm^3$

$c_{ref}$      Reference liquid phase $O_2$ concentration, $mol/cm^3$

$D_{eff}$      Effective diffusion coefficient of $O_2$ in a flooded agglomerate particle, $cm^2/s$

$D_{ON}^0$      Binary diffusion coefficient of $O_2$ and $N_2$ in a free gas stream, $cm^2/s$

$D_{OW}^0$      Binary diffusion coefficient of $O_2$ and water vapor in a free gas stream, $cm^2/s$

$D_{NW}^0$      Binary diffusion coefficient of $N_2$ and water vapor in a free gas stream, $cm^2/s$

$E$      Equilibrium potential of a cathode in reference to a standard $H_2$ electrode, V

$E_O^0$      Standard potential of a cathode in reference to a standard $H_2$ electrode, V

$F$      Faraday's constant, 96487 C/eq

$F$      $F$ distribution

H        Henry's constant, $[\text{mol/cm}^3\,(l)]/[\text{mol/cm}^3\,(g)]$

I        Steady state operating current density, $\text{A/cm}^2$

**I**        Identity matrix

$i_{\text{ref}}$        Exchange current density of the $O_2$ reduction reaction evaluated a reference $O_2$ concentration of $1.0 \times 10^{-6}$ $\text{mol/cm}^3$ in a flooded agglomerate particle, $\text{A/cm}^3$

**J**        The matrix of the partial derivatives of the dependent variable with respect to estimation parameters evaluated at all the experimental data point.

$j_O$        Steady state generation rate of $O_2$ gas per unit volume of the cathode CAL, $\text{mol/cm}^3$

$l_B$        Thickness of the GDL, cm

$l_c$        Thickness of the CAL, cm

P        Total gas pressure, atm

R        Universal gas constant, 8.3143 J/mol/K

**R**        Correlation matrix

$R_a$        Radius of an agglomerate particle, cm

$S^2$        Squared residual

$S_E$        Unbiased estimate of the variance

$S_{x,\theta_j}$        Sensitivity coefficient, $\partial x / \partial \theta_j$

$S_{\eta,\theta_j}$        Sensitivity coefficient, $\partial \eta / \partial \theta_j$

$t$        Student's $t$ distribution

T        Absolute temperature, K

$x$        Steady state mole fraction of $O_2$ in the gas pores

$z$        Normalized spatial coordinate in either the GDL or CAL, $0 \le z \le 1$

*w*     Mole fraction of water vapor in the gas pores

Greek symbols

$\boldsymbol{\theta}$     Parameter vector to be estimated

$\theta_j^*$     Point estimate of parameter $\theta_j$

$\eta$     Steady state over-potential, V

$\varphi_B$     Volume fraction of gas pores in the GDL

$\varphi_c$     Volume fraction of gas pores in the CAL

$\kappa_{eff}$     Effective ionic conductivity of the electrolyte, S/cm

$\Phi_1$     Steady state cathode potential, V

$\Phi_1^*$     Experimental steady state cathode potential, V

Subscripts

B     GDL

c     CAL

T     Transpose

-1     Inverse

# List of Figures

Fig. 6       The distribution of the dimensionless $O_2$ reduction current in the catalyst

layer of a low-pressure PEMFC air cathode (P=1.3 atm) with the change of the operating current density.

Fig. 7       Comparison of the simulated polarization curves of a high-pressure air

cathode (P=5.1 atm) and three low-pressure $O_2$ cathodes (P=1.3 atm).

Unless otherwise indicated on a plot, the point estimates obtained from the simultaneous fit were assigned to all the parameters in the simulations.

Fig. 8       Comparison of the simulated polarization curves of a low-pressure

PEMFC air cathode (P=1.3 atm). Except for the parameter values indicated on a plot, the point estimates obtained from the simultaneous fit were assigned to all the remaining parameters in the simulations.

Table I Parameters used for the steady state polarization model of a PEMFC cathode operated at 70 °C

| Parameter | Value | Comments |
|---|---|---|
| $D_{ON}^0$ | 0.230 cm$^2$/s | Ref. 20 (T=316 K, P=1 atm) [*] |
| $D_{OW}^0$ | 0.282 cm$^2$/s | Ref. 20 (T=308 K, P=1 atm) [*] |
| $D_{NW}^0$ | 0.293 cm$^2$/s | Ref. 20 (T=298 K, P=1 atm) [*] |
| $l_B$ | 0.04 cm | Measured on E-TEK GDL |
| $l_c$ | 0.0015 cm | Measured |
| $b$ | 0.0261 V [**] | Refs. 15-19 |
| H | 0.0277 | Ref. 18 |
| | [mol/cm$^3$($l$)]/[mol/cm$^3$($g$)] | |
| $E_O^0$ | 1.20 V | Ref. 18 |

[*] $D_{ij}^0(T,P) = D_{ij}^0(T_1,P_1) \times \dfrac{P_1}{P} \times \left(\dfrac{T}{T_1}\right)^{1.8}$

[**] A value on a $\Phi_1$ *vs.* ln(-I) plot

Table II Comparison of the 95% confidence intervals estimated from the simultaneous fit to three experimental polarization curves with those estimated from the independent fits

| | Simultaneous fit | Independent fit (P=1.3 atm) | Independent fit (P=2.3 atm) | Independent fit (P=3.3 atm) |
|---|---|---|---|---|
| $\varphi_B$ | $0.1991 \pm 6.676 \times 10^{-4}$ | $0.2013 \pm 2.521 \times 10^{-3}$ | $0.1980 \pm 1.019 \times 10^{-3}$ | $0.1966 \pm 6.341 \times 10^{-4}$ |
| $\varphi_c$ | $(3.933 \pm 0.2578) \times 10^{-2}$ | $(3.366 \pm 0.3669) \times 10^{-2}$ | $(3.925 \pm 0.6124) \times 10^{-2}$ | $(4.216 \pm 0.7155) \times 10^{-2}$ |
| $i_{ref}$ (A/cm$^3$) | $(7.198 \pm 0.8226) \times 10^{-4}$ | $(1.036 \pm 0.1829) \times 10^{-3}$ | $(6.408 \pm 1.409) \times 10^{-4}$ | $(5.152 \pm 1.081) \times 10^{-4}$ |
| $D_{eff}/R_a^2$ (s$^{-1}$) | *$(3.052 \pm 1.637) \times 10^{3}$ | $(8.173 \pm 16.46) \times 10^{3}$ | $(2.226 \pm 2.605) \times 10^{3}$ | $(1.534 \pm 1.694) \times 10^{3}$ |
| $\kappa_{eff}$ ($\Omega$/cm) | $(9.947 \pm 1.004) \times 10^{-3}$ | $(7.750 \pm 2.230) \times 10^{-3}$ | $(1.207 \pm 0.2822) \times 10^{-2}$ | $(1.468 \pm 0.3385) \times 10^{-2}$ |
| $S_E$ (V) | $1.239 \times 10^{-2}$ | $0.8916 \times 10^{-2}$ | $1.010 \times 10^{-2}$ | $0.9766 \times 10^{-2}$ |

*If the value of $D_{eff}$ is assumed to be $2.199 \times 10^{-6}$ cm$^2$/s,[8] the value of R$_a$ is found to be in the range of $0.2165 \leq R_a < 0.3942$ μm, which is generally consistent with the values reported in refs.11 and12.

Table III    Comparison of the computer time required by a personal computer with an 866 MHz CPU for the calculation of nonlinear model equations

|  | With banded Jacobian matrix (user-supplied) | With banded Jacobian matrix (not user-supplied ) | With dense Jacobian matrix (not user-supplied) |
|---|---|---|---|
| Calculating 200 nonlinear model equations 6 times | 1.27 s | 2.07 s | 31.3 s |
| Calculating 1200 nonlinear model equations once | 1.64 s | 7.35 s | 188 s |
| Numerical efficiency summary | Good | Fair | Poor |

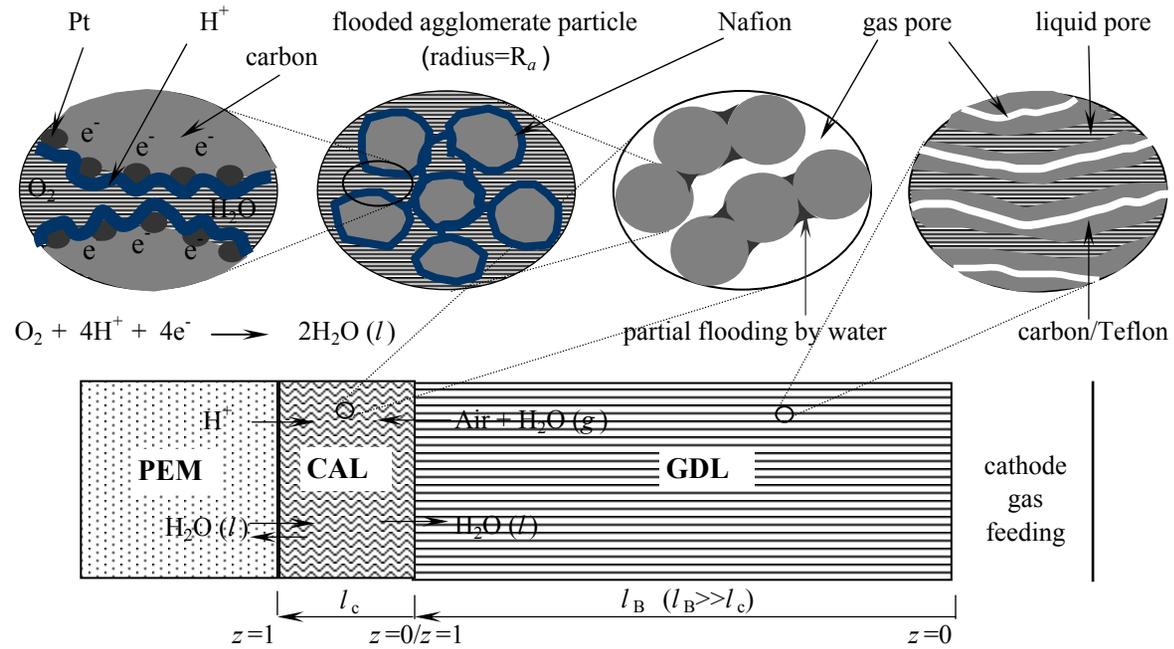

Fig. 1 Q. Guo *et al.*

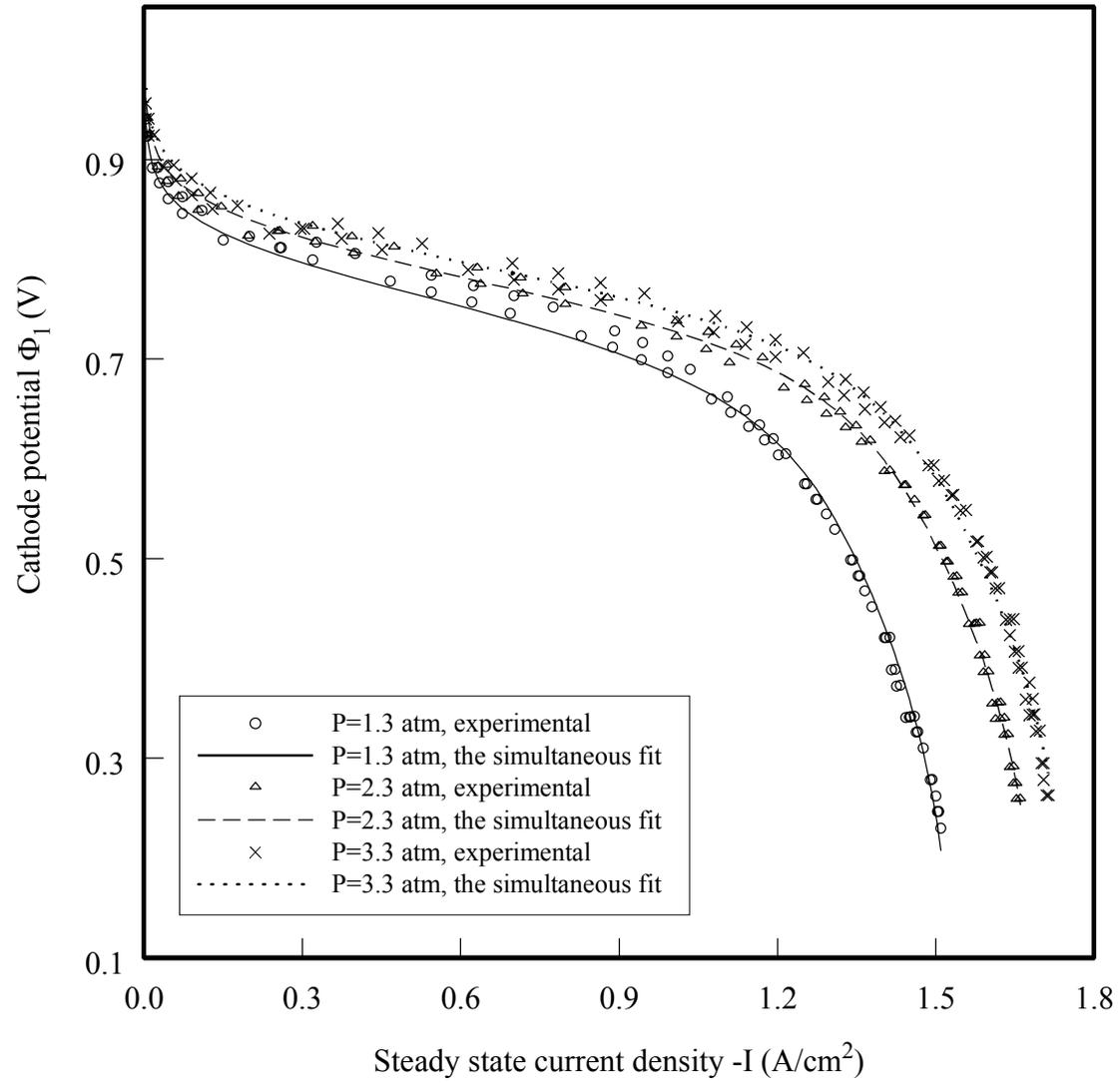

Fig. 2-1 Q. Guo *et al.*

Fig. 2-2 Q. Guo *et al.*

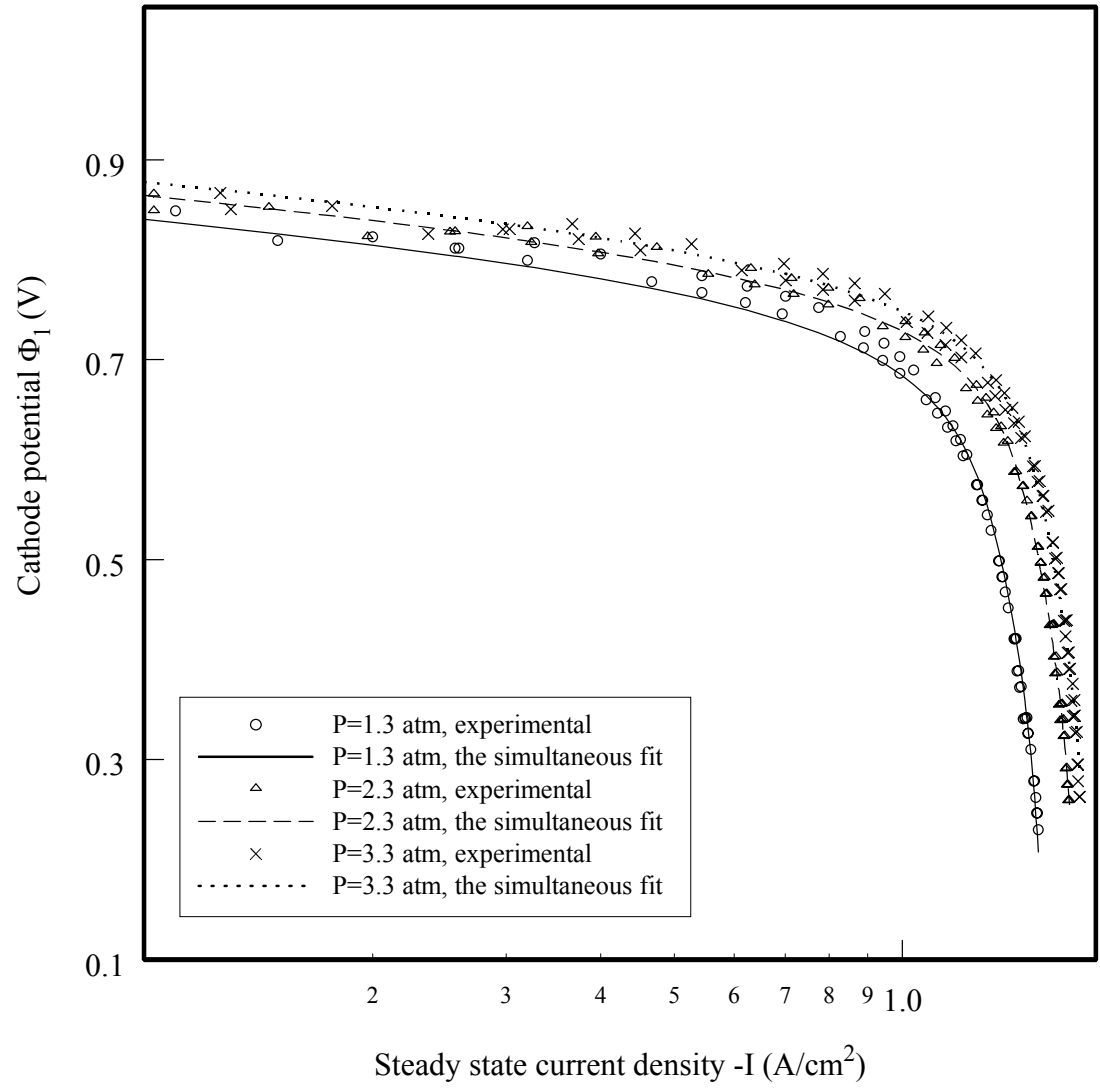

Fig. 3 Q. Guo *et al.*

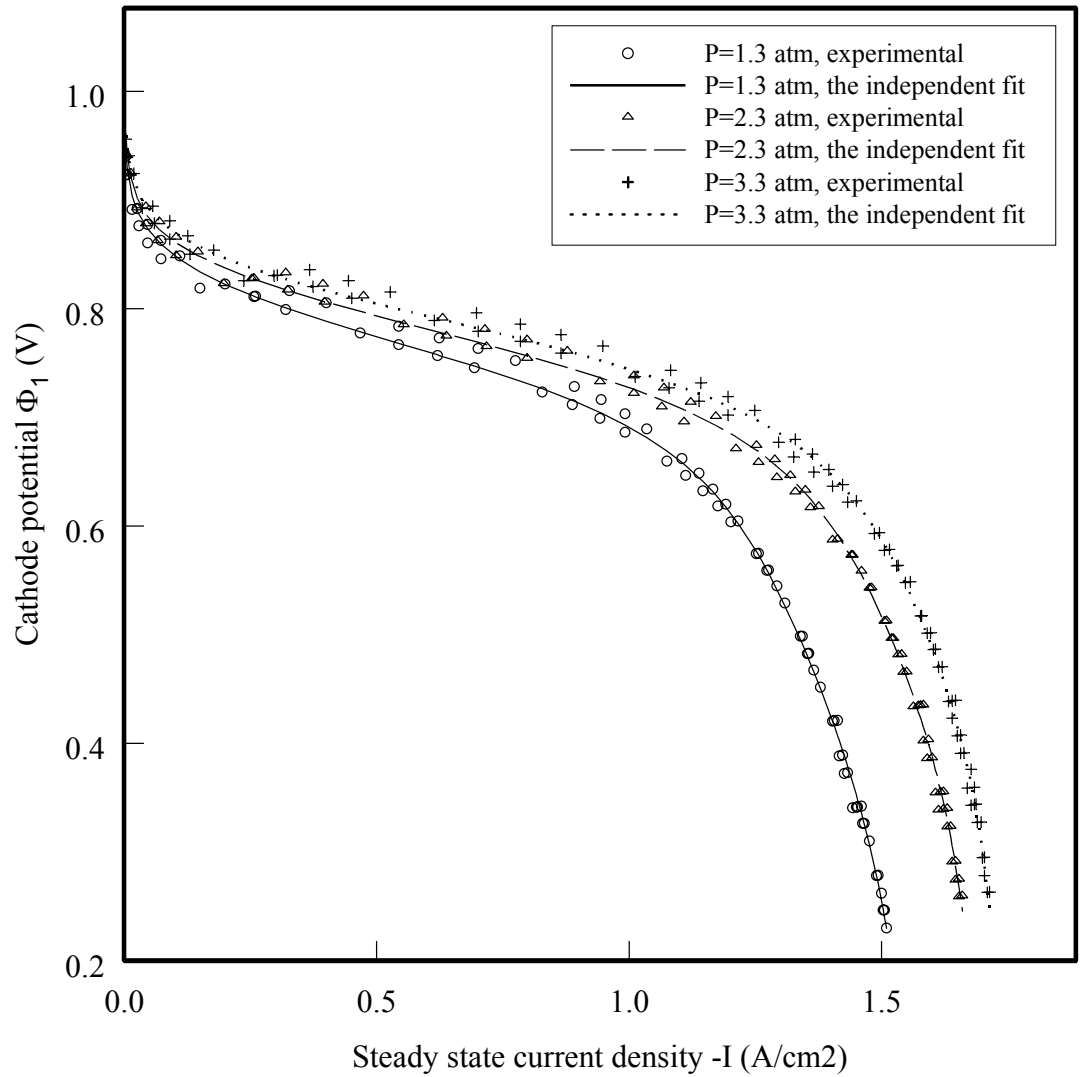

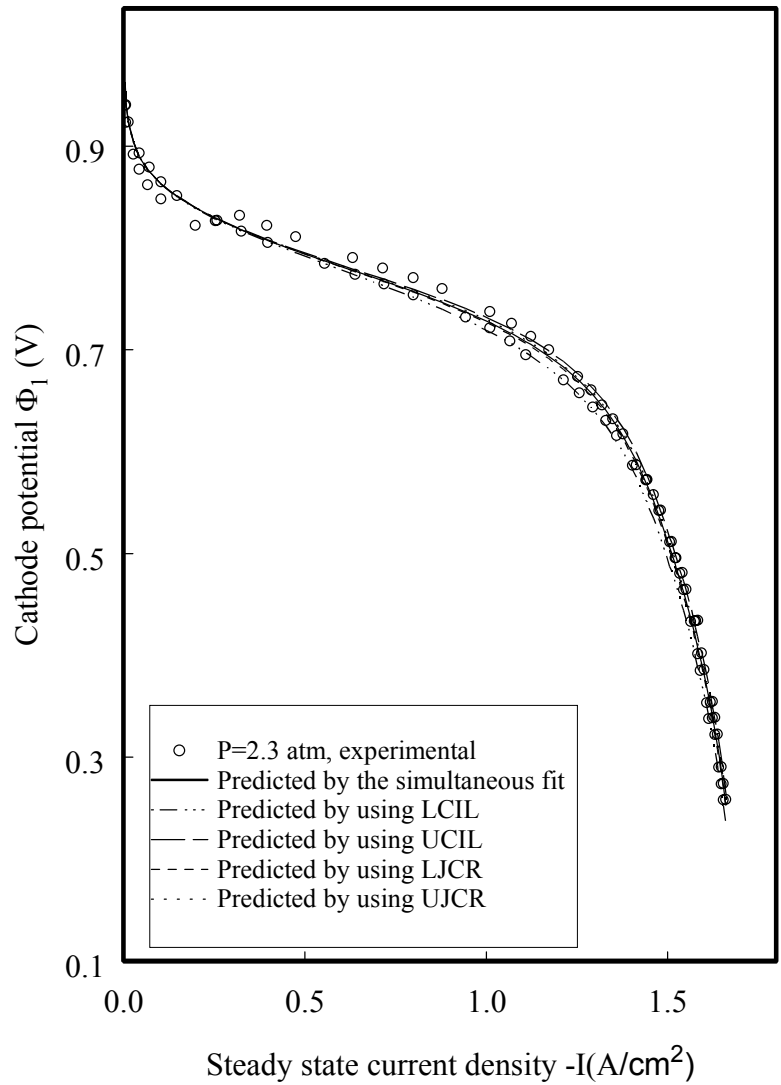

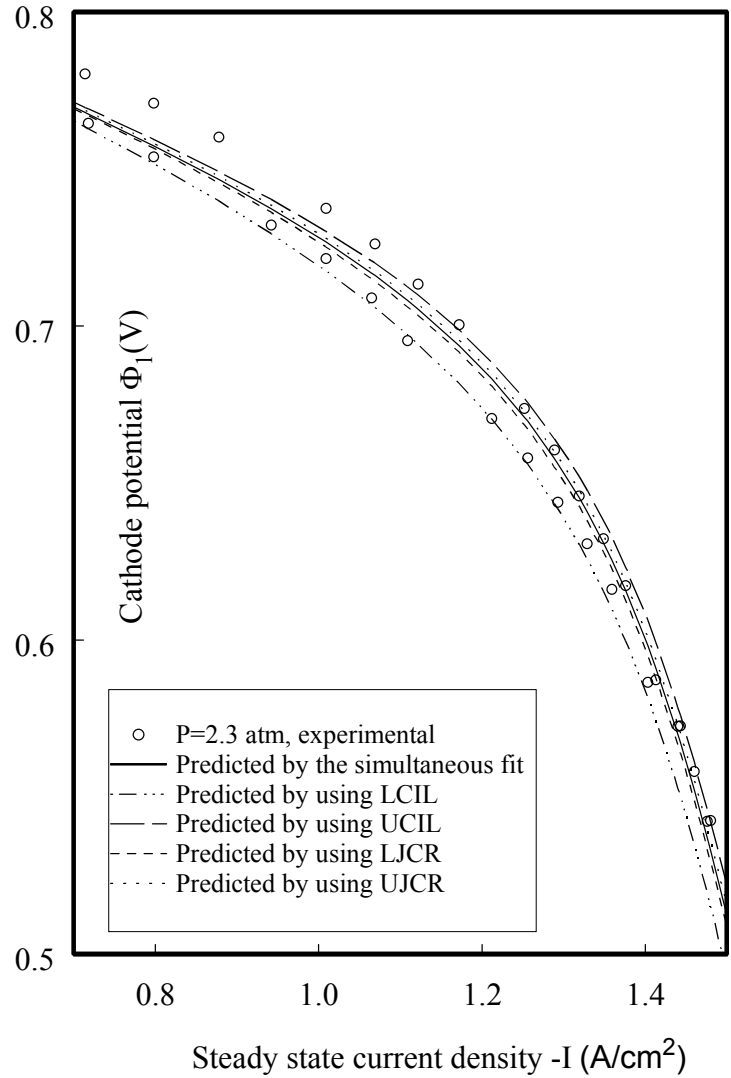

Fig. 5 Q. Guo *et al.*

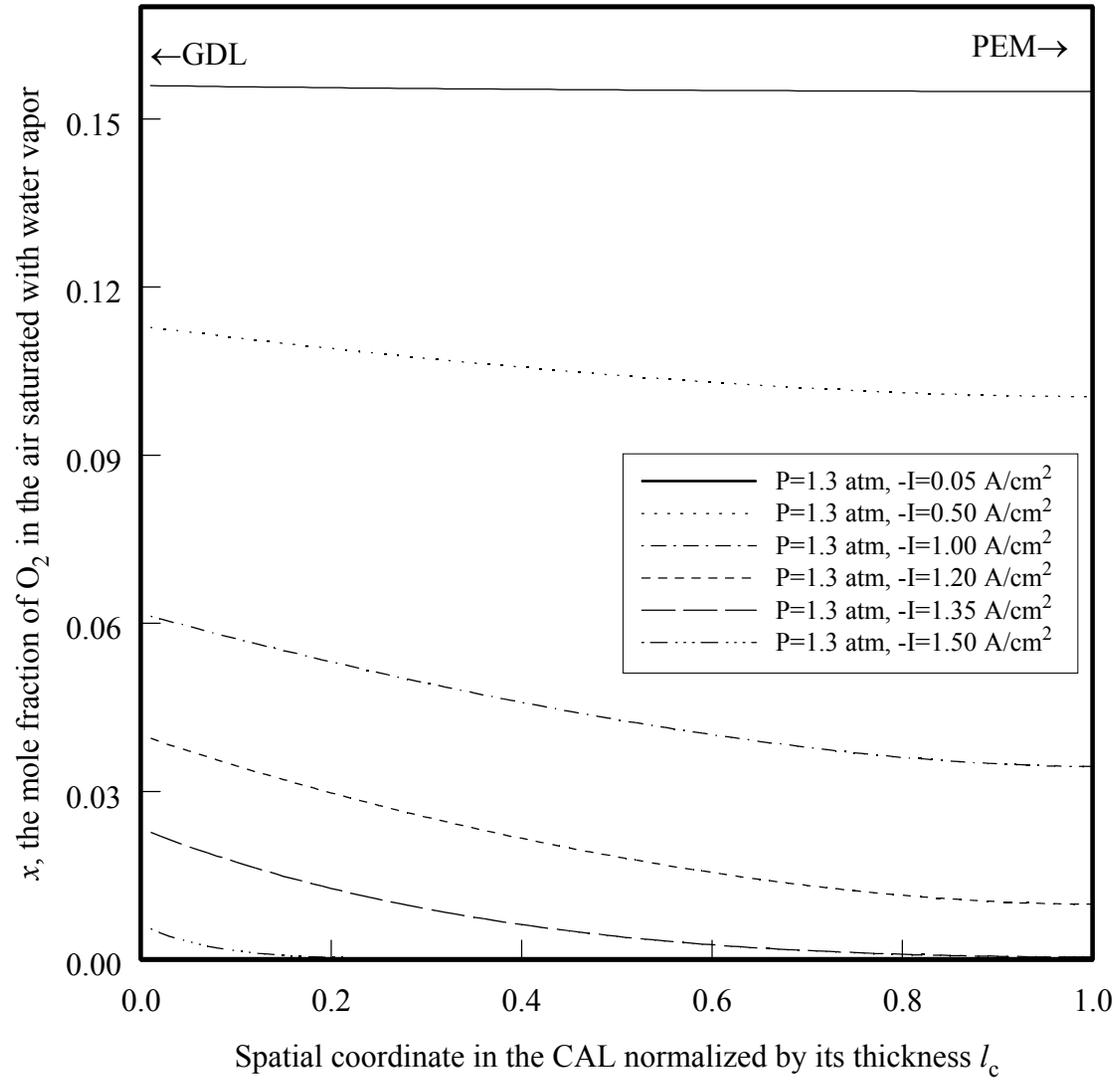

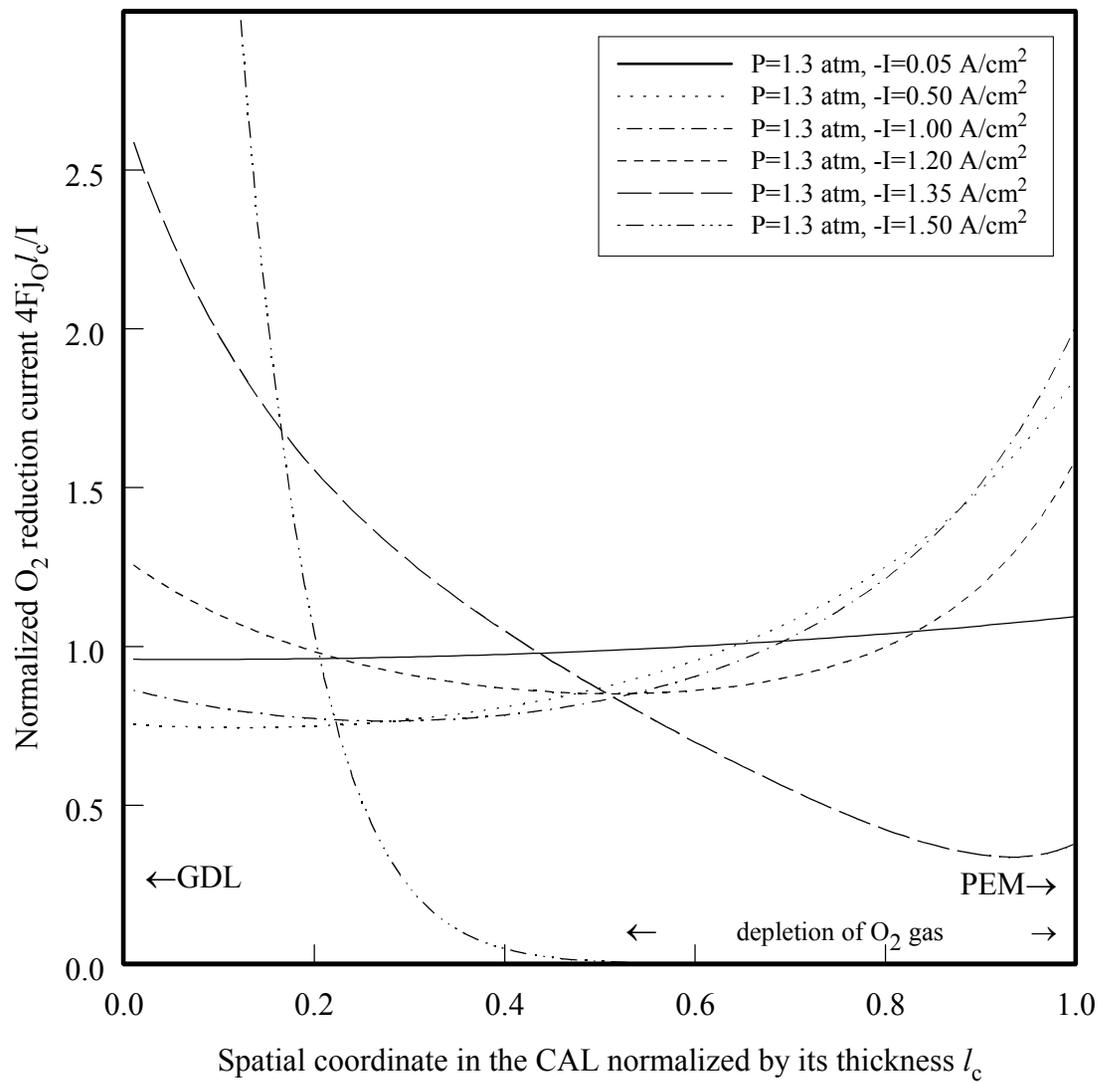

Fig. 7 Q. Guo *et al.*

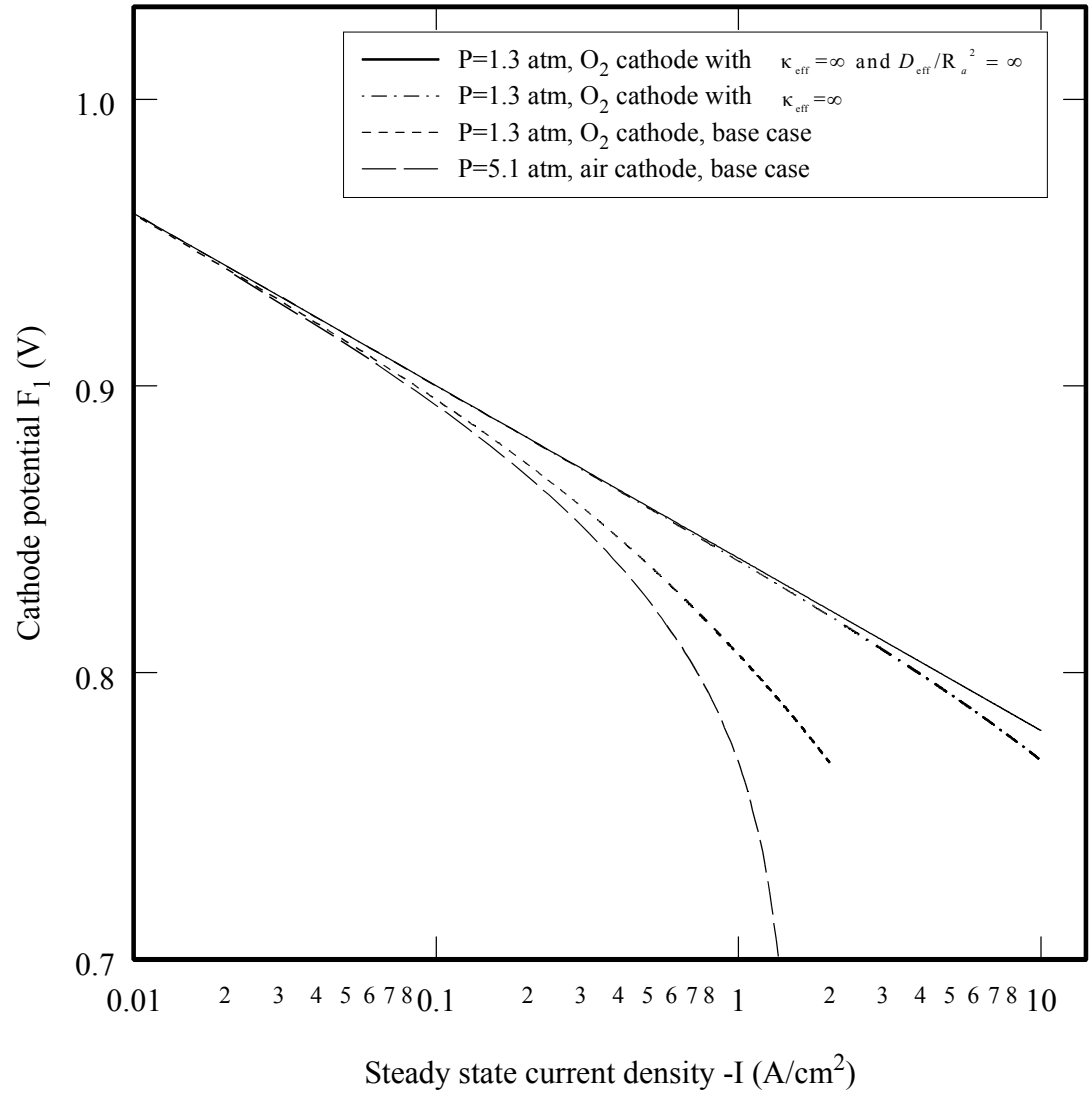

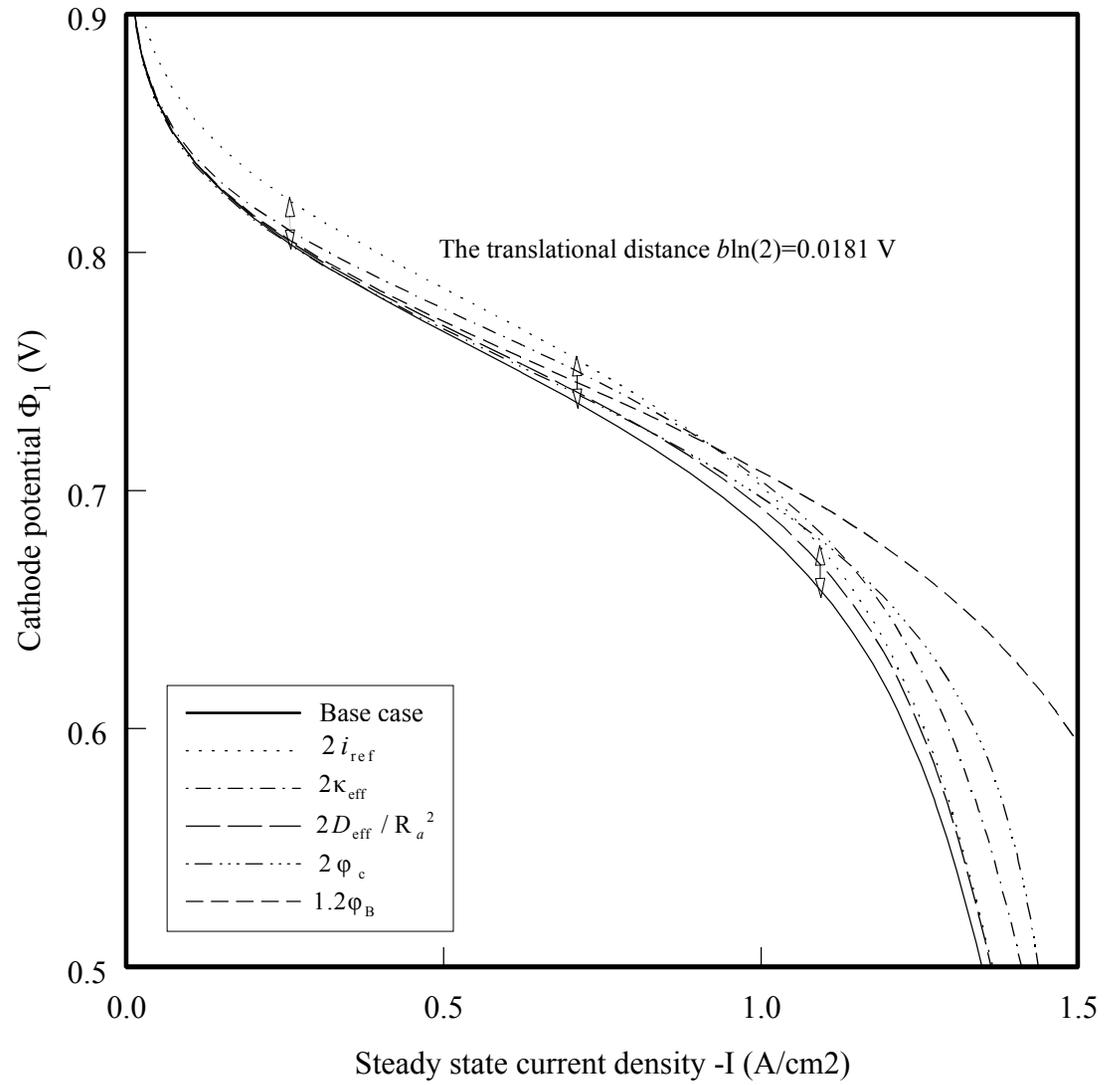